\documentclass[aip, jap, reprint, superscriptaddress]{revtex4-1}

\emergencystretch=\maxdimen
\usepackage[english]{babel}
\usepackage{amsmath,amsthm}
\usepackage{amsfonts}
\usepackage{xspace}
\usepackage{bm}
\usepackage{booktabs}
\usepackage{graphicx}
\usepackage{dcolumn}
\usepackage[mathlines]{lineno}
\usepackage[colorlinks]{hyperref}  
\usepackage{gensymb}
\usepackage{epstopdf}
\usepackage{newfloat}
\usepackage[usenames, dvipsnames]{color}
\DeclareFloatingEnvironment[name={Figure S}]{suppfigure}
\DeclareFloatingEnvironment[name={S}]{suppEquation}

\newcommand{\muB}{\ensuremath{\mu_\mathrm{B}}\xspace}

\newcommand{\aR}{\ensuremath{\alpha_\mathrm{R}}\xspace}

\newcommand{\tFM}{\ensuremath{t_\mathrm{FM}}\xspace}
\newcommand{\Aex}{\ensuremath{A_\mathrm{ex}}\xspace}
\newcommand{\Ms}{\ensuremath{M_\mathrm{s}}\xspace}
\newcommand{\Kcub}{\ensuremath{K_\mathrm{c}}\xspace}
\newcommand{\Jex}{\ensuremath{J_\mathrm{ex}}\xspace}

\newcommand{\By}{\ensuremath{B_\mathrm{y}}\xspace}
\newcommand{\Bsw}{\ensuremath{B_\mathrm{sw}}\xspace}

\newcommand{\Bthso}{\ensuremath{B^\mathrm{th}_\mathrm{so}}\xspace}
\newcommand{\Bso}{\ensuremath{B_\mathrm{so}}\xspace}
\newcommand{\Jth}{\ensuremath{J^\mathrm{th}}\xspace}

\newcommand{\BsoVecTop}{\ensuremath{\bm{B^\mathrm{top}_\mathrm{so}}}\xspace}
\newcommand{\BsoVecBot}{\ensuremath{\bm{B^\mathrm{bot}_\mathrm{so}}}\xspace}
\newcommand{\MagVecTop}{\ensuremath{\bm{m^\mathrm{top}}}\xspace}
\newcommand{\MagVecBot}{\ensuremath{\bm{m^\mathrm{bot}}}\xspace}
\newcommand{\TotMagVec}{\ensuremath{\bm{m}}\xspace}
\newcommand{\NeelVec}{\ensuremath{\bm{l}}\xspace}

\newcommand{\mx}{\ensuremath{m_\mathrm{x}}\xspace}
\newcommand{\my}{\ensuremath{m_\mathrm{y}}\xspace}
\newcommand{\mz}{\ensuremath{m_\mathrm{z}}\xspace}
\newcommand{\mxy}{\ensuremath{m_\mathrm{x,y}}\xspace}

\newcommand{\mxtop}{\ensuremath{m^\mathrm{top}_\mathrm{x}}\xspace}
\newcommand{\mytop}{\ensuremath{m^\mathrm{top}_\mathrm{y}}\xspace}

\newcommand{\mxbot}{\ensuremath{m^\mathrm{bot}_\mathrm{x}}\xspace}
\newcommand{\mybot}{\ensuremath{m^\mathrm{bot}_\mathrm{y}}\xspace}

\newcommand{\lx}{\ensuremath{l_\mathrm{x}}\xspace}
\newcommand{\ly}{\ensuremath{l_\mathrm{y}}\xspace}
\newcommand{\lz}{\ensuremath{l_\mathrm{z}}\xspace}

\newcommand{\xUnit}{\ensuremath{\bm{\hat{x}}}\xspace}
\newcommand{\yUnit}{\ensuremath{\bm{\hat{y}}}\xspace}

\begin{document}

\title{Switching of biaxial synthetic antiferromagnets: a micromagentic study}%

\author{Michael S. Ackermann}%
\affiliation{ 
Academy of Integrated Science, Virginia Tech, Blacksburg, VA 24061, USA
}%
\affiliation{ 
Department of Mathematics, Virginia Tech, Blacksburg, VA 24061, USA
}%
\author{Satoru Emori}%
\email{
semori@vt.edu
}
\affiliation{ 
Department of Physics, Virginia Tech, Blacksburg, VA 24061, USA
}%

\date{November 21, 2018}

\begin{abstract}
We simulate the switching behavior of nanoscale synthetic antiferromagnets (SAFs), inspired by recent experimental progress in spin-orbit-torque switching of crystal antiferromagnets.
The SAF consists of two ferromagnetic thin films with in-plane biaxial anisotropy and interlayer exchange coupling.
Staggered field-like Rashba spin-orbit torques from the opposite surfaces of the SAF induce a canted net magnetization, which triggers an orthogonal torque that drives 90$^\circ$ switching of the N\'eel vector.  
Such dynamics driven by the field-like spin-orbit torque allows for faster switching with increased Gilbert damping, without a significant detrimental increase of the threshold switching current density. 
Our results point to the potential of SAFs as model systems, based on simple ferromagnetic metals, to mimic antiferromagnetic device physics.
\end{abstract}
\maketitle

\section{Introduction}
Antiferromagnets are considered promising material platforms for ultrafast spintronic information-technology devices that are highly stable against external magnetic fields~\cite{Jungwirth2016, Jungwirth2018, Baltz2018}. 
Recent experimental studies have demonstrated switching of the antiferromagnetic order (N\'eel vector) between two orthogonal states in epitaxial antiferromagnetic conductors of CuMnAs~\cite{Wadley2016a} and Mn$_2$Au~\cite{Bodnar2018}. This N\'eel switching is driven by a current-induced ``field-like'' spin-orbit torque (SOT) that acts locally in opposite directions on the two magnetic sublattices of the antiferromagnet~\cite{Zelezny2014, Zelezny2018}. 
The key ingredient for this non-vanishing field-like N\'eel SOT is the inversion asymmetry around each magnetic atom (i.e., Mn) that is intrinsic to the specific crystal structure of CuMnAs and Mn$_2$Au. 
However, the synthesis of epitaxial CuMnAs and Mn$_2$Au with the correct crystal structure may not be straightforward, and so far no other conductive collinear antiferromagnets with the compatible crystal structure for the field-like N\'eel SOT have been realized~\cite{Wadley2016a, Bodnar2018}. 
It has also been shown that epitaxial antiferromagnetic insulator NiO can be switched by a SOT from an adjacent metal with strong spin-orbit coupling (e.g., Pt)~\cite{Chen2018a, Moriyama2018}. In this case, the limitation may be the relatively small magnetoresistance signal (i.e., spin-Hall magnetoresistance~\cite{Fischer2018}) to read out the N\'eel vector state. 
Furthermore, it is generally difficult to apply conventional laboratory-based characterization techniques (e.g., magnetometry, ferromagnetic resonance, magnetic microscopy, etc.) to study the fundamental properties of antiferromagnets. These points above may constitute a serious obstacle to studying and engineering viable materials for antiferromagnetic spintronics. 

\begin{figure}[b]
  \includegraphics [width=1.00\columnwidth] {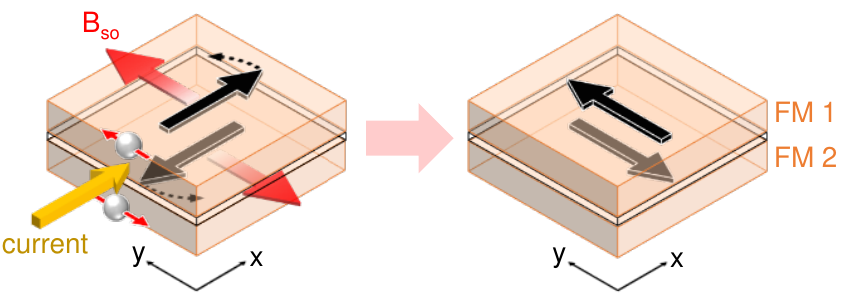}
  \centering
  \caption{\label{fig:cartoon}Schematic of the synthetic antiferromagnet (SAF) consisting of two ferromagnetic metal (FM) layers. The SAF has two orthogonal easy axes along the $x$- and $y$-axes. The current-induced spin-orbit field \Bso switches the antiferromagnetic order (N\'eel vector) from the $x$- to $y$-axis.}
\end{figure}

Here, we study by micromagnetic simulations the switching behavior of synthetic antiferromagnets (SAFs)~\cite{Duine2018} as a model system analogous to intrinsic crystal antiferromagnets. The SAFs consist of two in-plane biaxial ferromagnetic metals (FMs) whose magnetizations are locked antiparallel to each other by interlayer exchange coupling (e.g., through the Ruderman-Kittel-Kasuya-Yosida mechanism across a non-ferromagnetic metal such as Cr or Ru)~\cite{Grunberg1986, Parkin1990}.  
Such biaxial FMs can be readily synthesized by epitaxial growth on a cubic single-crystal substrate, e.g., body-centered-cubic Fe on MgO (001) or GaAs (001)~\cite{Mattson1995, Postava1997, Bowen2001, Chen2016c, Lee2017}. 
This SAF structure has two orthogonal easy axes in the film plane defined by cubic magnetocrystalline anisotropy.
These two digital states, represented by orthogonal N\'eel vector orientations, can be read through anisotropic magnetoresistance~\cite{Wadley2016a, Bodnar2018}; e.g., when the N\'eel vector is oriented parallel (transverse) to the sense current, the SAF exhibits a higher (lower) electrical resistance~\cite{Marti2014,Moriyama2015}. 
Switching is achieved when opposite local fields are applied orthogonal to the magnetizations of the two FM layers (Fig.~\ref{fig:cartoon}), analogous to the opposite local fields applied to the two sublattices in CuMnAu and Mn$_2$Au~\cite{Wadley2016a, Bodnar2018}. 
In the SAF, the required symmetry breaking for such opposite local fields occurs at the layer interfaces. We simulate the effect of the interfacial Rashba spin-orbit fields (field-like SOTs)~\cite{Gambardella2011a, Fan2013, Emori2016} arising from the top and bottom surfaces of the SAF interfaced with,  e.g., an oxide capping layer and substrate~\cite{Xu2012, Ibrahim2016}. 
Our study therefore suggests a possible pathway for simple SAF spintronic devices that inherit some of the switching behavior of antiferromagnets.

Advantages of SAFs have been reported previously for engineering stable pinned and free layers in spin valves~\cite{Smith2008, Bandiera2010, Devolder2012}, rapid motion of domain walls~\cite{Saarikoski2014, Yang2015a, Komine2016, Lepadatu2017}, and SOT-driven switching of perpendicular magnetization~\cite{Shi2017, Bi2017, Zhang2018}.  
In contrast with these prior devices based on 180$^\circ$ switching, we emphasize that our proposed approach is based on \emph{90$^\circ$} switching.  
To the best of our knowledge, our study is the first to numerically examine such orthogonal switching in SAFs, specifically driven by field-like SOT. 
The orthogonal orientation between the initial magnetization and the spin-orbit field in each FM layer (Fig.~\ref{fig:cartoon}) maximizes the torque on the magnetization and enables rapid switching. This switching scheme driven by the field-like torque also allows for faster switching by increasing the Gilbert damping parameter without an adverse increase of the threshold switching current density. Our simulations indicate that SAFs with realistic material parameters are robust against up to $\sim$1~T of external magnetic field and can be switched in $\sim$0.1~ns at a reasonable current density of $\lesssim$$10^{11}$~A/m$^2$. 
We also note that this proposed device scheme is operated by two orthogonal current lines, analogous to four-terminal toggle magnetic random access memories (MRAMs)~\cite{Engel2005}. The use of the field-like SOT as proposed here, instead of Oersted fields in toggle MRAMs, may enable alternative scalable memory devices.

\section{Model Parameters}
Magnetic switching was simulated using the Mumax3 micromagnetics package~\cite{Vansteenkiste2014}. 
A series of square samples with different widths of 26 to 400 nm were studied with a lateral cell size of 2 or 4 nm. 
Each FM layer had the following fixed properties: thickness \tFM = 1.5 nm, exchange constant \Aex = 20 pJ/m, and the cubic anisotropy constant \Kcub = 30 kJ/m$^3$ with the easy axes parallel to the square edges. 
In most simulations, we set the saturation magnetization \Ms at 1700 kA/m (typical value for Fe), the Gilbert damping parameter $\alpha$ at 0.01 (typical value for nanometer-thick FMs), and the interlayer exchange coupling energy density \Jex at $-$0.2 or $-$1 mJ/m$^2$ (where the negative sign indicates antiferromagnetic interlayer coupling). We note that the values of \Jex used here are similar to those experimentally achieved in SAFs consisting of FMs~\cite{Parkin1990, Fullerton1993, Lepadatu2017, Khodadadi2017a}. 

\begin{figure}[tb]
  \includegraphics [width=0.92\columnwidth] {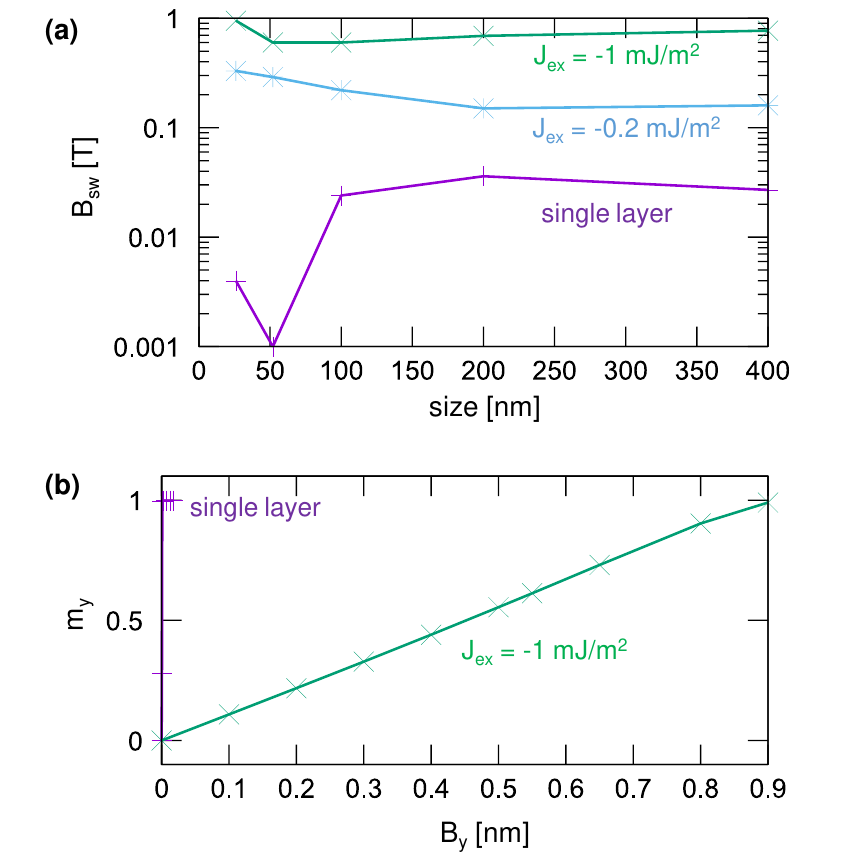}
  \centering
  \caption{\label{fig:stability}(a) External magnetic field \Bsw required to switch the magnetization from the $x$-axis to $y$-axis for samples of different lateral sizes. (b) Equilibrium magnetization component \my along the $y$-axis versus external magnetic field \By. Here the lateral sample size is 52 nm.}
\end{figure}

\section{Results and Discussion}
\label{sec:results}
\subsection{Stability against global magnetic field}
\label{subsec:stability}
We first compare the stability of the magnetization state against a global external field in the SAFs with their single-layer FM counterparts. 
With the initial magnetizations set parallel to the $x$-axis (i.e., $\mxtop = 1$, $\mxbot = -1$, and $\mytop = \mybot = 0$), an external magnetic field \By along the $+y$-direction was applied. The critical switching field \Bsw is defined as \By required to pull the total magnetization, $\TotMagVec = \frac{1}{2}(\MagVecTop + \MagVecBot)$, to the $y$-direction, i.e., $\my > 0.99$. Figure~\ref{fig:stability}(a) shows that the single-layer FMs switch at low values of $\Bsw~\lesssim~0.01$ T, indicating that these samples are vulnerable to spontaneous switching from external stray fields.  
As evidenced by the substantial variation in \Bsw\ -- as much as an order of magnitude -- with lateral size, the switching behavior of the single-layer FMs is also heavily impacted by the device geometry, e.g., due to dipolar fields from the sample edges. 
A slight variation in the shape or edge defects of single-layer in-plane FM devices can lead to a random distribution of switching thresholds. 

The SAFs show about an order of magnitude greater \Bsw than the single-layer FMs. As shown in Fig.~\ref{fig:stability}(a), \Bsw is enhanced with increasing \Jex. For $\Jex = -1$~mJ/m$^2$ readily achievable in realistic SAFs~\cite{Parkin1990, Fullerton1993, Lepadatu2017, Khodadadi2017a}, an external field of nearly 1 T is required to orient the magnetization along the $y$-direction. As shown in Fig.~\ref{fig:stability}(b), while the single-layer FM undergoes abrupt switching at low \By, the SAF undergoes a gradual magnetization rotation until the magnetizations of the two layers are fully oriented along the field direction. 
We also note that \Bsw only varies by a factor of $\approx$2 with the lateral sample dimensions of the SAFs (Fig.~\ref{fig:stability}(a)), indicating that the dipolar fields from the sample edges play relatively little role. 
The SAFs are therefore shown to be significantly more stable against disturbances from external magnetic fields, and this stability is largely independent of the sample geometry. 
We emphasize that the stability at fields of $\sim$0.1-1 T can be achieved in SAFs consisting of simple FMs (e.g., Fe), in contrast with intrinsic crystal antiferromagnetic compounds~\cite{Wadley2016a, Bodnar2018} for which epitaxial growth is more challenging. 

\subsection{Threshold spin-orbit field for switching}
\label{subsec:current}
Having demonstrated the stability of the SAFs, we compute how much spin-orbit field is required to switch the antiferromagnetic order in the SAFs between the $x$- and $y$-axes (e.g., Fig.~\ref{fig:cartoon}). 
For example, the magnetization of the top (bottom) FM layer, initially oriented along the +$x$-direction ($-x$-direction), sees an effective current-induced field pointing along the -$y$-direction ($+$$y$-direction). When the magnitude of this effective field is sufficiently large, the magnetization overcomes the cubic anisotropy energy barrier and switches from the $x$-axis to $y$-axis. 
Unlike a global magnetic field (Sec.~\ref{subsec:stability}) that cants the magnetizations toward the parallel state and hence results in a large interlayer antiferromagnetic exchange energy penalty, the local spin-orbit field rotates the magnetization of each layer while maintaining the mostly antiparallel magnetization alignment across the layers. 
We discuss the details of the switching process in Sec.~\ref{subsec:mechanism}.

We define the threshold \Bthso as the effective local field required to switch the N\'eel vector, $\NeelVec = \frac{1}{2}(\MagVecTop - \MagVecBot )$, to the $y$-axis, i.e., $|\ly| > 0.99$.
We simulated two cases where (1) only the top layer sees the spin-orbit field (and the bottom layer magnetization is dragged by the top layer magnetization), and (2) the top and bottom layers see the spin-orbit field in opposite directions (Fig.~\ref{fig:cartoon}). These two configurations of the spin-orbit field would arise by enabling an interfacial Rashba field-like SOT at (1) only the top surface of the SAF and (2) both the top and bottom surfaces of the SAF. 

\begin{figure}[tb]
  \includegraphics [width=0.92\columnwidth] {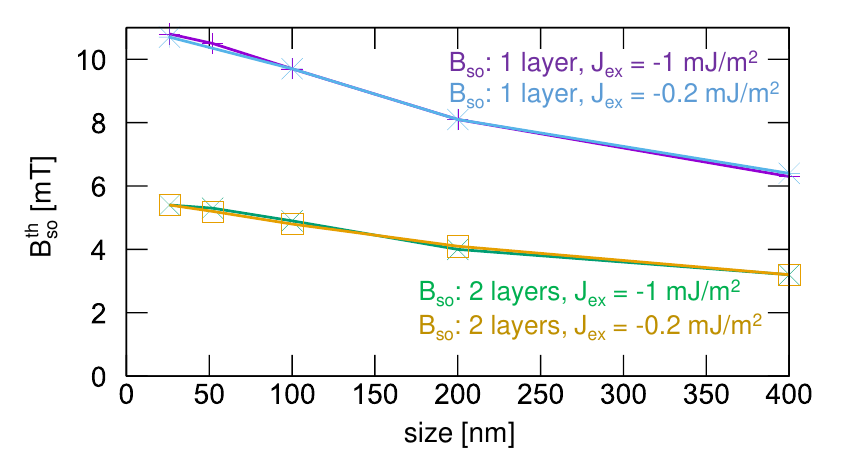}
  \centering
  \caption{\label{fig:current}Threshold current-induced spin-orbit field \Bthso required to switch the N\'eel vector from the $x$-axis to $y$-axis for SAF samples with different lateral dimensions.}
\end{figure}

Figure~\ref{fig:current} plots the computed \Bthso against the SAF lateral size. \Bthso is somewhat dependent on the lateral sample size, increasing by nearly a factor of 2 when the lateral sample size is decreased from 400 to 26 nm, as the mode of switching transitions from incoherent to coherent. 
More importantly, we find a factor of 2 reduction in \Bthso with the current-induced field active at both the top and bottom surfaces compared to just one. This finding confirms that the spin-orbit field is additive and that engineering the Rashba effect at both surfaces would lead to a more efficient SAF device. 
It should also be noted that $|\Jex|$ does not affect \Bthso, suggesting that biaxial SAFs can be switched efficiently regardless of the strength of interlayer exchange coupling. 
Here, since the cubic magnetic anisotropy energy density $\Kcub \sim 10^4$ J/m$^3$ is significantly smaller than the interlayer exchange energy density $|\Jex|/\tFM \sim 10^5-10^6$ J/m$^3$,  the energy barrier for 90$^\circ$ switching of the N\'eel vector is mostly determined by \Kcub rather than $|\Jex|$. This finding is consistent with a prior study of 180$^\circ$ switching  in SAFs, where the energy barrier is governed by uniaxial magnetic anisotropy~\cite{Devolder2012}. However, we show in the next subsection (Sec.~\ref{subsec:mechanism}) that the interlayer exchange coupling can influence the switching speed by generating a torque on the N\'eel vector.

Provided that the spin-orbit field arises entirely from the interfacial Rashba-Edelstein effect, we can estimate the critical threshold current density for switching \Jth from \Bthso with~\cite{Manchon2008, Kim2012a}
\begin{equation}
\Jth = \frac{\muB\Bthso\Ms}{\aR P},
\end{equation}
where \muB is the Bohr magneton, \aR is the Rashba parameter, and $P$ is the effective spin polarization (proportional to the exchange interaction between the Rashba-induced spin accumulation and the FM magnetization). For \Jth to be comparable to $\lesssim$$10^{11}$~A/m$^2$ recently reported in antiferromagnetic memory prototypes~\cite{Wadley2016a, Bodnar2018, Chen2018a, Moriyama2018}, the product $\aR P$ would need to be $\gtrsim$$0.1$ eV$\cdot$\AA.  This is reasonably achieved with Rashba parameters similar to those reported in oxide systems~\cite{Varignon2018, Caviglia2010, Santander-Syro2014, Lesne2016, Wang2017e, Xu2012, Ibrahim2016}. 
While the field-like SOT has not received as much attention (compared to the damping-like SOT) for FM-based device applications, an enhanced interfacial Rashba spin-orbit field would be a robust driving force to efficiently switch a biaxial SAF memory.

\subsection{Time-dependence of switching}

\begin{figure}[tb]
  \includegraphics [width=0.92\columnwidth] {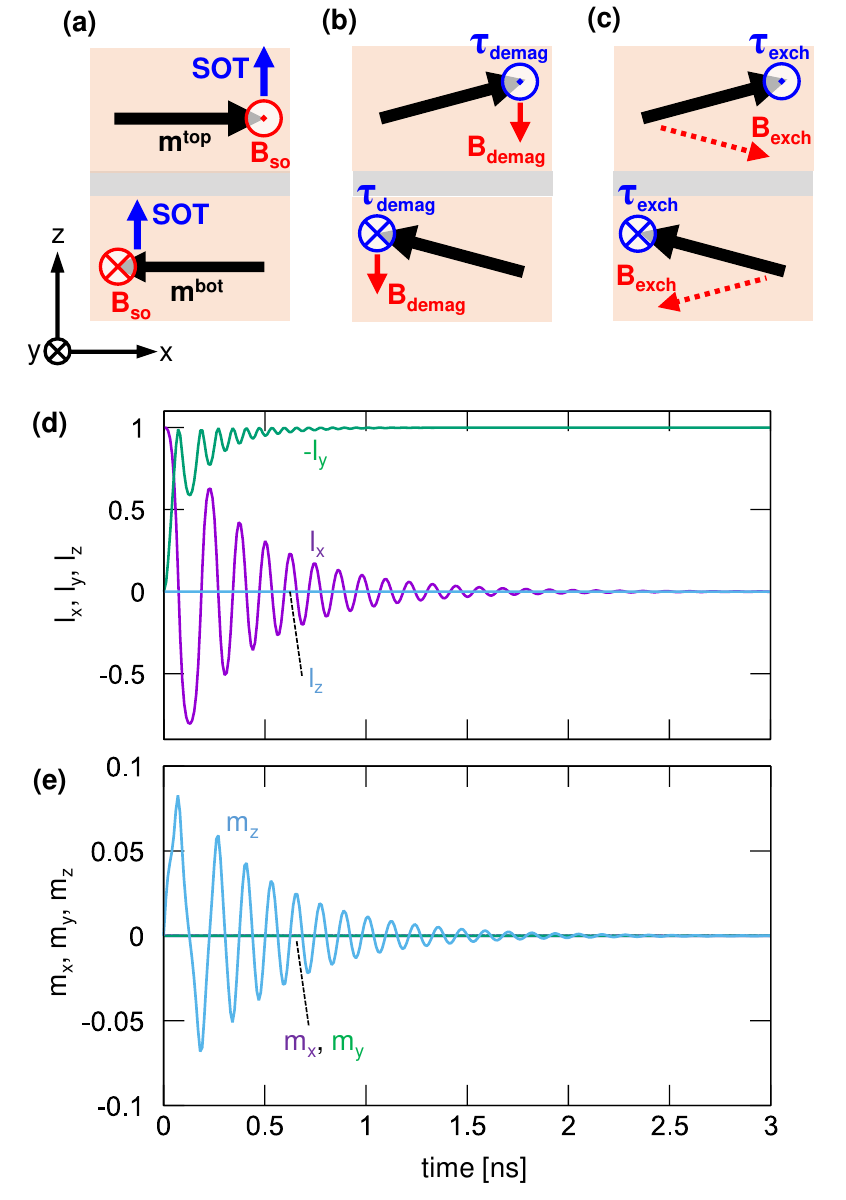}
  \centering
  \caption{\label{fig:mech}(a,b,c) Schematics of the torques due to the (a) spin-orbit field \Bso, (b) demagnetizing field \ensuremath{B_\mathrm{demag}}, and (c) interlayer antiferromagnetic exchange field \ensuremath{B_\mathrm{exch}}.  (d,e) Time traces of the (d) N\'eel vector (\lx, \ly, \lz) and (e) total magnetization (\mx, \my, \mz) at \Bso = 6 mT in the 52-nm-wide SAF sample with \Ms = 1700 kA/m, \Jex = -1 mJ/m$^2$, $\alpha$ = 0.01.}
\end{figure}

\label{subsec:mechanism}
Finally, we discuss the mechanism and time dependence of SOT-driven switching in the SAFs. In the following, switching is driven by opposite local spin-obit fields acting on the top and bottom layers. The initial orthogonal configuration between the spin-orbit field (e.g., $\BsoVecTop||-\yUnit$, $\BsoVecBot||+\yUnit$) and the magnetization ($\MagVecTop||+\xUnit$, $\MagVecBot||-\xUnit$) in each layer maximizes the torque that initiates the switching process. When this SOT ($-|\gamma|\MagVecTop\times\BsoVecTop$, $-|\gamma|\MagVecBot\times\BsoVecBot$) is turned on, the magnetization is tilted out of the film plane in the \emph{same} direction in \emph{both} layers (Fig.~\ref{fig:mech}(a)), thereby giving rise to a finite $z$-component in the total magnetization \mz. This out-of-plane canting then yields two torques along $-\yUnit$ in the top layer ($+\yUnit$ in the bottom layer): (1) a torque due to the out-of-plane demagnetizing field (Fig.~\ref{fig:mech}(b)) and (2) a torque due to the interlayer antiferromagnetic exchange penalty (Fig.~\ref{fig:mech}(c)). These demagnetizing and antiferromagnetic-exchange torques have the same symmetry to drive 90$^\circ$ switching of the N\'eel vector \NeelVec from the $x$-axis to the $y$-axis.

\begin{figure*}[tb]
  \includegraphics [width=1.90\columnwidth] {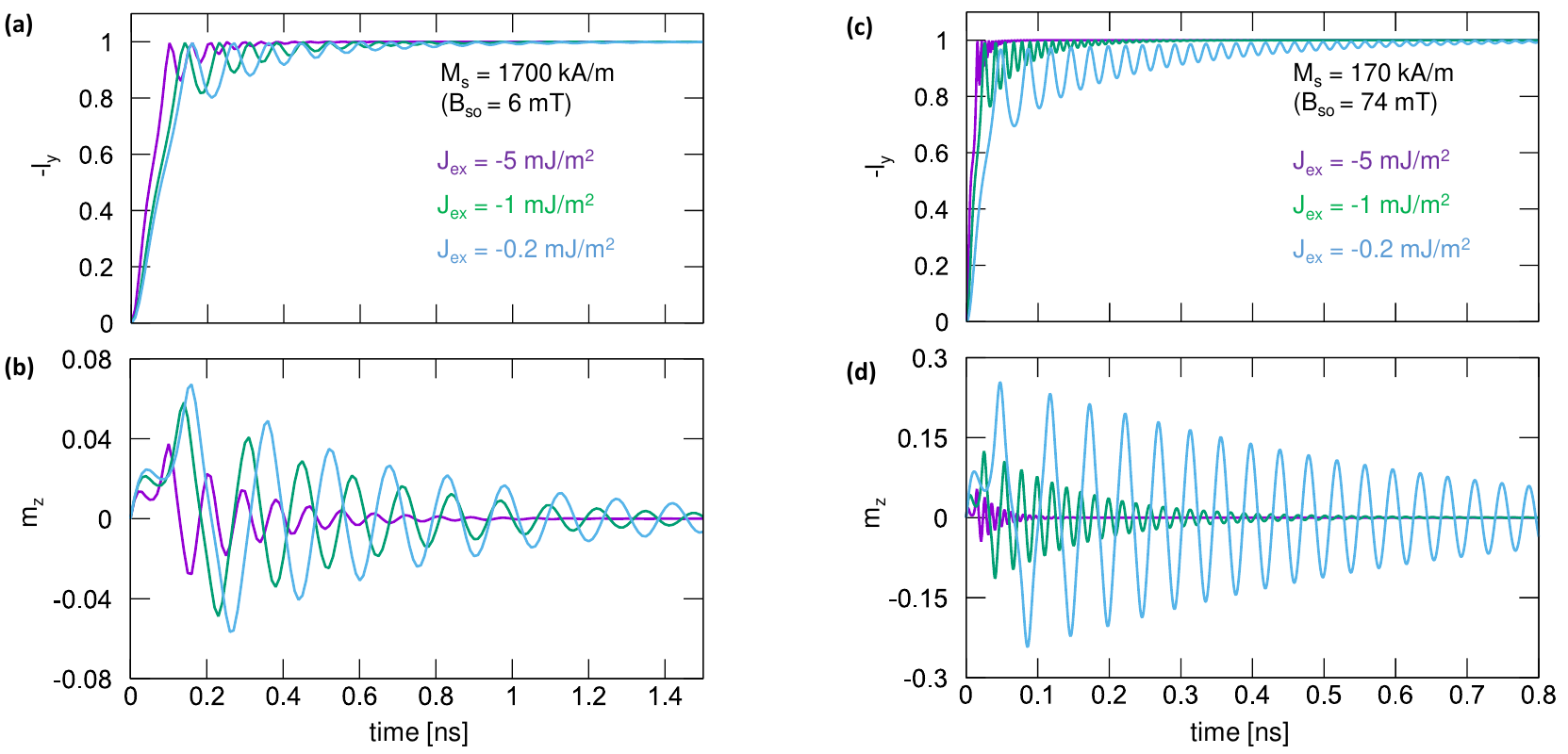}
  \centering
  \caption{\label{fig:Ms}Time traces of the (a,c) N\'eel vector component \ly and (b,d) total magnetization component \mz for samples with (a,b) \Ms = 1700 kA/m and (c,d) \Ms = 170 kA/m at different strengths of interlayer exchange coupling \Jex. The magnitude of the spin-orbit field \Bso is chosen to be slightly above the threshold for switching. The sample width is 52 nm and $\alpha = 0.01$.}
\end{figure*}

An exemplary time evolution of the N\'eel vector and total magnetization is shown in Fig.~\ref{fig:mech}(d,e). The initial rise in $\mz$ confirms the out-of-plane tilting of the magnetization, while $\mxy \approx 0$ indicates that the in-plane magnetization components remain compensated between the two layers. Moreover, the damped oscillation of \NeelVec (Fig.~\ref{fig:mech}(d)) exhibits a phase offset of $\pi/2$ with respect to \mz (Fig.~\ref{fig:mech}(e)), i.e., the time rate of change of \NeelVec is maximized when $|\mz|$ exhibits a maximum. This relation confirms that the torque on \NeelVec is indeed related to the magnetization canting $|\mz|$.

The relative contributions of the torques can be tuned by varying the saturation magnetization \Ms, since a smaller value of \Ms should decrease the demagnetizing torque contribution. Figure~\ref{fig:Ms} compares the time-dependence of switching for SAFs with \Ms = 1700 kA/m and 170 kA/m, each with different strengths of interlayer exchange coupling \Jex. For each \Ms, the magnitude of the spin-orbit field \Bso is chosen to be slightly above the threshold for switching \Bthso. In the case of \Ms = 1700 kA/m, the switching speed changes only by a factor of $\approx$2 when $\Jex$ is varied by a factor of 25 (Fig.~\ref{fig:Ms}(a,b)). Evidently, for SAFs consisting of high-moment FMs (e.g., Fe), the out-of-plane demagnetizing torque dominates the switching process, whereas the antiferromagnetic-exchange torque plays a relatively minor role. We thus find that although SAFs have zero net magnetization at equilibrium, their dynamics can be driven predominantly by the demagnetizing field from nonequilibrium magnetization. By contrast, in the case of \Ms = 170 kA/m, increasing $|\Jex|$ results in an order of magnitude faster switching (Fig.~\ref{fig:Ms}(c,d)), indicating that the antiferromagnetic-exchange torque plays a relatively major role when the constituent FMs have low magnetization.

\begin{figure}[h!]
  \includegraphics [width=0.92\columnwidth] {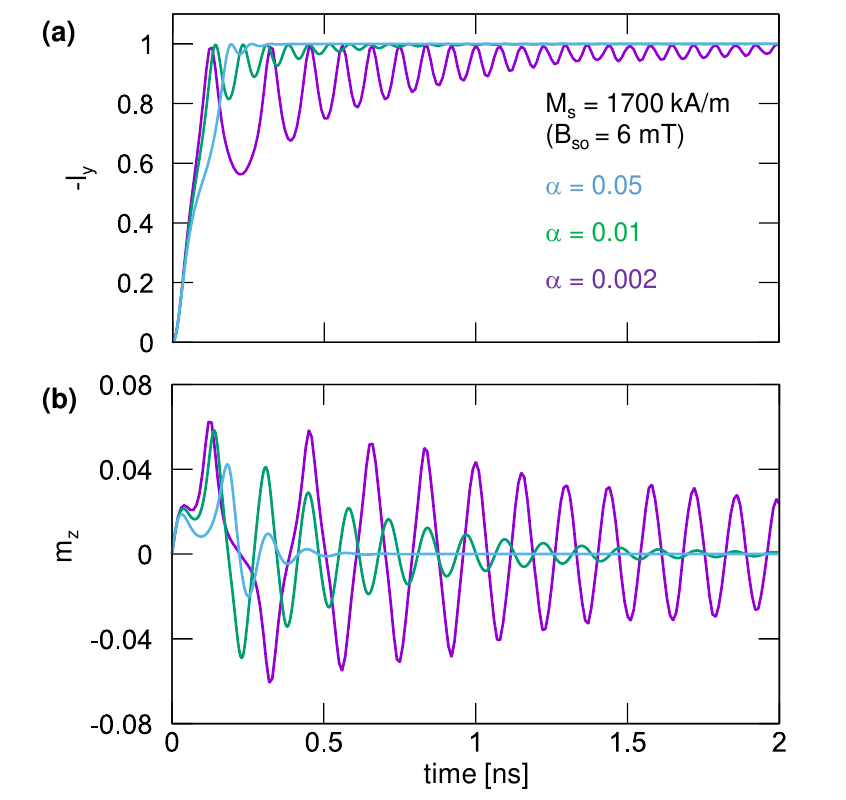}
  \centering
  \caption{\label{fig:damping}Time traces of the (a) N\'eel vector component \ly and (b) total magnetization component \mz at \Bso = 6 mT in the 52-nm-wide SAF sample for with \Ms = 1700 kA/m and \Jex = -1 mJ/m$^2$ at different Gilbert damping parameters $\alpha$.}
\end{figure}

Enhancing the interlayer exchange coupling is not necessarily an effective way to speed up switching in SAFs because (1) the exchange torque may not be the dominant driving mechanism for switching if the constituent FMs have high \Ms and,  (2) even if the exchange torque dominates, it would be practically difficult to increase $|\Jex|$ well above  $\sim$1~mJ/m$^2$. We therefore explore an alternative method of enhancing the switching speed by increasing the Gilbert damping parameter $\alpha$, which is experimentally more straightforward (e.g., through alloying the FM with a small concentration of rare-earth metal~\cite{Woltersdorf2009}).
The threshold current density for switching driven by the field-like torque is not significantly affected by damping~\footnote{We find that \Bthso increases by at most $\approx$30 \% when $\alpha$ is increased by a factor of 25 from 0.002 to 0.05.}.  
This is in contrast with coherent switching of a single-domain in-plane nanomagnet driven by a damping-like torque, where the threshold current density is inversely proportional to the damping parameter~\cite{Liu2012}; since lower damping would prolong the magnetization oscillations before settling along the equilibrium orientation, damping-like-torque-driven switching leads to a trade-off between reducing the power consumption (threshold switching current density) and the switching time. 
Our proposed scheme of utilizing the field-like torque in the biaxial SAF allows for speeding up switching by increasing the damping parameter, without adversely affecting the threshold switching current density. 

Figure~\ref{fig:damping} shows the influence of the damping parameter $\alpha$ on the time evolutions of \NeelVec and \TotMagVec in an SAF. 
The oscillations around the $y$-axis are significantly suppressed at higher values of $\alpha$ in Fig.~\ref{fig:damping}. 
We note, however, that by increasing $\alpha$ further to $\gtrsim0.1$, the switching process becomes overdamped and is hence slowed down. 
These results indicate that the switching time is minimized with a moderately large value of $\alpha$. 
The time traces shown in Fig.~\ref{fig:damping} are obtained at $\Bso = 6$ mT, which is only slightly above the threshold for switching for the simulated 52-nm-wide device (Fig.~\ref{fig:current}). The switching time can be decreased further with a greater spin-orbit field (current density). Our results thus suggest that 90$^\circ$ switching in an SAF device can be accomplished in $\lesssim 0.1$ ns at a reasonable current density of $\sim$$10^{11}$ A/m$^2$, provided a sufficiently strong interfacial Rashba-Edelstein effect (as discussed in Sec.~\ref{subsec:current}). While $\sim$0.1-ns switching has been demonstrated for SOT-driven perpendicular anisotropy memories, the required current density exceeds $10^{12}$ A/m$^2$~\cite{Garello2014}. Biaxial SAFs may therefore be an attractive power-efficient alternative to conventional spintronic memory platforms.

\section{Summary}
We have demonstrated by micromagnetic simulations that biaxial SAFs -- consisting of two antiferromagnetically-coupled FMs -- are stable against large external magnetic fields and can be switched efficiently with a field-like SOT. 
Even though SAFs have zero net magnetization at equilibrium, the field-like SOT yields a finite nonequilibrium magnetization, which gives rise to switching driven mostly by the torque from the demagnetizing field, particularly if the SAF consists of high-moment FMs (e.g., Fe). The 90$^\circ$ switching scheme can enable fast dynamics, especially when combined with moderately high Gilbert damping. 
Such SAFs can be readily engineered from simple FMs and are attractive model systems that mimic some of the  dynamics of intrinsic crystal antiferromagnets.

\vspace{1pt}
This work was supported in part by the Luther and Alice Hamlett Undergraduate Research Support Program in the Academy of Integrated Science at Virginia Tech. 
We thank Minh-Hai Nguyen for helpful discussion. 

\vspace{1pt}
Note added in proof: After submission of this manuscript, we became aware of an experimental report by Moriyama \textit{et al}.~\cite{Moriyama2018a} that demonstrates switching of antiferromagnetic order in amorphous CoGd synthetic antiferromagnets, driven by a damping-like SOT.

\end{document}